\documentclass[preprint]{aastex631}

\newcommand\Msun{$M_\odot$}

\shorttitle{Evolution of the TW Hya Disk Shadow}
\shortauthors{Debes et al.}
\begin{document}

\title{The surprising evolution of the shadow on the TW Hya disk\footnote{Based on observations made with the NASA/ESA Hubble Space Telescope, obtained at the Space Telescope Science Institute, which is operated by the Association of Universities for Research in Astronomy, Inc., under NASA contract NAS5-26555. These observations are associated with programs \#16228 and \#13573. }}

\correspondingauthor{John Debes}
\email{debes@stsci.edu}

\author[0000-0002-1783-8817]{John Debes}
\affiliation{AURA for ESA, Space Telescope Science Institute, 3700 San Martin Dr., Baltimore, MD 21218, USA}

\author[0000-0003-0856-679X]{Rebecca Nealon}
\affiliation{Department of Physics, University of Warwick, Coventry CV4 7AL, UK}
\affiliation{Centre for Exoplanets and Habitability, University of Warwick, Coventry CV4 7AL, UK}

\author[0000-0001-6410-2899]{Richard Alexander}
\affiliation{School of Physics and Astronomy, University of Leicester, University Rd., Leicester LE1 7RH, UK}

\author[0000-0001-6654-7859 ]{Alycia J. Weinberger}
\affiliation{Earth and Planets Laboratory, Carnegie Institution for Science, 5241 Broad Branch Rd NW, Washington, DC 20015}

\author[0000-0002-9977-8255]{Schuyler Grace Wolff}
\affiliation{Department of Astronomy, The University of Arizona, 933 N Cherry Ave, Tucson, AZ, 85750, USA}

\author[0000-0003-4653-6161]{Dean Hines}
\affiliation{Space Telescope Science Institute, 3700 San Martin Dr., Baltimore, MD 21218, USA}

\author[0000-0002-3138-8250]{Joel Kastner}
\affiliation{School of Physics \& Astronomy, Rochester Institute of Technology, 1 Lomb Memorial Dr., Rochester, NY 14623, USA}

\author[0000-0002-7639-1322]{Hannah Jang-Condell}
\affiliation{NASA Headquarters, 300 Hidden Figures Way SW, Washington, DC 20546, USA}

\author[0000-0001-5907-5179]{Christophe Pinte}
\affiliation{School of Physics and Astronomy, Monash University, Clayton Vic 8300, Australia}
\affiliation{Univ. Grenoble Alpes, CNRS, IPAG, 38000 Grenoble, France}

\author[0000-0002-8864-1667]{Peter Plavchan}
\affiliation{Department of Physics \& Astronomy, George Mason University, 4400 University Dr., Fairfax, VA 22030}

\author[0000-0003-3818-408X]{Laurent Pueyo}
\affiliation{Space Telescope Science Institute, 3700 San Martin Dr., Baltimore, MD 21218, USA}



\begin{abstract}

We report new total intensity visible light high contrast imaging of the TW Hya disk taken with the Space Telescope Imaging Spectrograph (STIS) on the Hubble Space Telescope (HST). This represents the first published images of the disk with STIS since 2016, when a moving shadow on the disk surface was reported. We continue to see the shadow moving in a counter-clockwise fashion, but in these new images the shadow has evolved into two separate shadows, implying a change in behavior for the occulting structure. Based on radiative transfer models of optically thick disk structures casting shadows, we infer that a plausible explanation for the change is that there are now two misaligned components of the inner disk. The first of these disks is located between 5-6au with an inclination of 5.5\arcdeg and PA of 170\arcdeg, the second between 6-7au with and inclination of 7\arcdeg and PA of 50\arcdeg. Finally, we speculate on the implications of the new shadow structure and determine that additional observations are needed to disentangle the nature of TW Hya's inner disk architecture.  

\end{abstract}



\section{Introduction} \label{sec:intro}

TW Hya is a nearby, well-studied face-on protoplanetary disk \citep[d=60~pc;i$\sim$5\arcdeg][]{edr3,qi04,huang18}. It has been observed from X-ray to radio wavelengths. Estimates of TW Hya's stellar mass generally lie in the range 0.6-0.8~M$_{\odot}$, with recent (dynamical) estimates favoring the upper end of this range \citep[e.g.][]{teague22}. Given its age ($\sim$10 Myr), this makes the TW Hya disk an interesting test case for planet formation in a large, relatively isolated, gas-rich disk orbiting a star somewhat less massive than the young Sun, but more massive than transiting young M dwarf systems such as AU Mic. \citep{teague19, sokal18}. HST has imaged the disk numerous times, including with WFPC2 \citep[IWA$\approx$0.7\arcsec;][]{krist00}, NICMOS Near-IR (NIR) F110W and F160W \citep[IWA$\approx$0.4\arcsec;][]{Weinberger:2002}, and STIS visible light imaging/spectroscopy \citep[IWA$\approx$0.2\arcsec;][hereafter D17]{Roberge:2005,debes17}. The disk is polarized, allowing visible and NIR images \citep{a15,boekel17,poteet18}. In the sub-mm, dust continuum emission extends from 3-$\sim$100~au\citep{andrews16,huang18,ilee22}, with multiple gaps or rings present that are close in radius to structures seen in scattered light and consistent with active planet formation \citep[][and references therein]{jangcondell12,andrews16,ueda20,macias20}. CO gas emission extends to $\sim$200~au, and scattered light from small dust extends as far out as 452~au\citep[][D17]{huang18}. 

The TW~Hya disk has an azimuthal surface brightness asymmetry between 30-150~au \citep{Roberge:2005,debes13}. The asymmetry has been attributed to inclination effects for flared disks \citep{debes13}, the forward scattering properties of dust \citep{Roberge:2005}, or the presence of a warp \citep{Roberge:2005,rosenfeld12}. Images of the asymmetry with STIS and NICMOS over six epochs between 0.6-2$\mu m$\ show that the asymmetry is not stationary; the apparent changes in disk illumination can be modeled with a rotating shadow with a constant angular velocity of 22.7$^\circ$/yr (P=15.9~yr). Assuming the shadow arises from an obstructing feature moving with Keplerian velocity constrains the location of the originating structure to $<5.6(\frac{M_\star}{0.7M_\odot})$~au (D17). ALMA CO maps of the outer disk also show the shadow \citep{teague22}. The position of the shadow in CO taken in 2019 is smaller than expected based on the predictions of D17. 

The TW~Hya shadow is one of a handful of other disks that show evidence of inner occulting structures, often interpreted as mis-aligned inner disks casting narrow or broad shadows depending on the inclination difference with the outer disk \citep[e.g. PDS 66, HD 34700, HD 16192, HD 142527, HD 139614, and RXJ1604.3-2130A][]{wolff,uyama20,bertrang18,marino15,muro-arena20,pinilla18}. Some of these disks also show evidence for  time variability, but robustly sampling the variability timescales is difficult for high contrast imaging targets. The robustness of the shadow detection and coverage of its characteristic timescale of motion make the shadow on TW~Hya's disk one of the best characterized shadows to date that may be caused by active planet formation.

\citet{nealon19} investigated the behavior of shadows due to misaligned disks that precess and warps associated with planetary companions orbiting the host star, in order to understand TW Hya's shadow. They found that massive planets with inclinations relative to the outer disk of a few degrees or more were sufficient to create shadows of the right amplitude in the outer disk either through a warp exterior to the planet's orbit or from a misaligned disk interior to the planet. This work motivated a multi-Cycle HST program with the STIS coronagraph in order to disentangle these two origins. TW Hya has only been observed for about half the observed periodicity, and a misaligned, precessing disk will continue to show asymmetries in the disk for a whole precession period, while features associated with a misaligned planet will disappear for half an orbital period. It is possible that both structures are present and contributing to the shadow behavior.

We report here the first epoch results from our monitoring program of TW Hya's disk to further track the evolution of the asymmetry and putative shadow. In \S \ref{sec:obs} we describe the new STIS observations. In \S \ref{sec:az} we report our findings of the latest epoch which show a marked departure in the behavior of the shadow. In \S \ref{sec:implications} we discuss the possible explanation of this evolution. Finally we present our Conclusions in \S \ref{sec:conclusions}.

\begin{figure}
    \centering
    \plotone{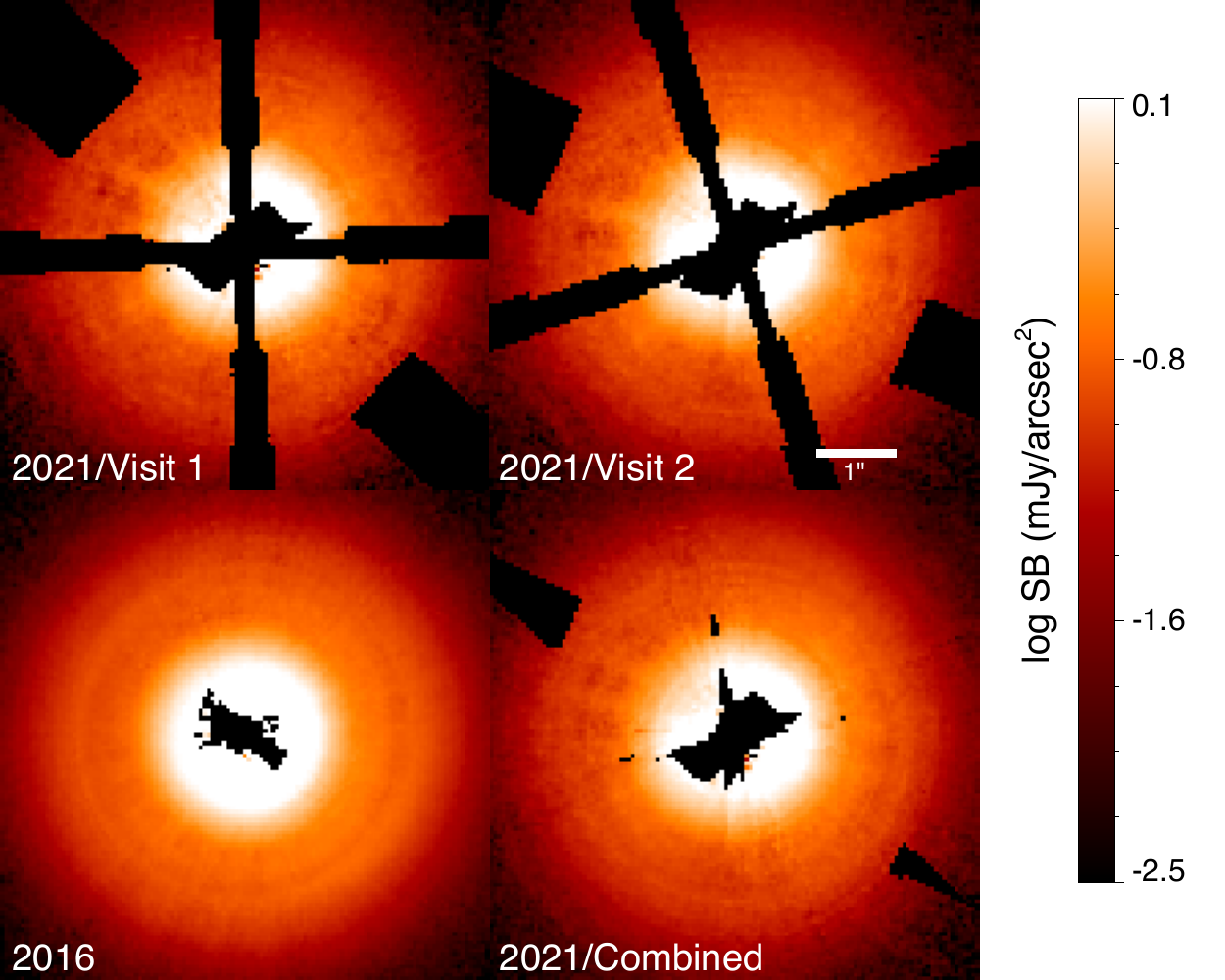}
    \caption{Logarithmic images of TW Hya from various epochs. North is up and East to the left of the images. (upper left) Image of the first visit of Program 16228 taken on 07 June 2021. Black areas represent missing data due to diffraction spikes and occulting masks. (upper right) The second visit of Program 16228, taken at a different spacecraft orientation. (lower left) Fully combined images of TW Hya from 2016 (lower right) Combination of Visit 1 and Visit 2 from Program 16228.}
    \label{fig:f1}
\end{figure}

\begin{figure}
    \centering
    \plotone{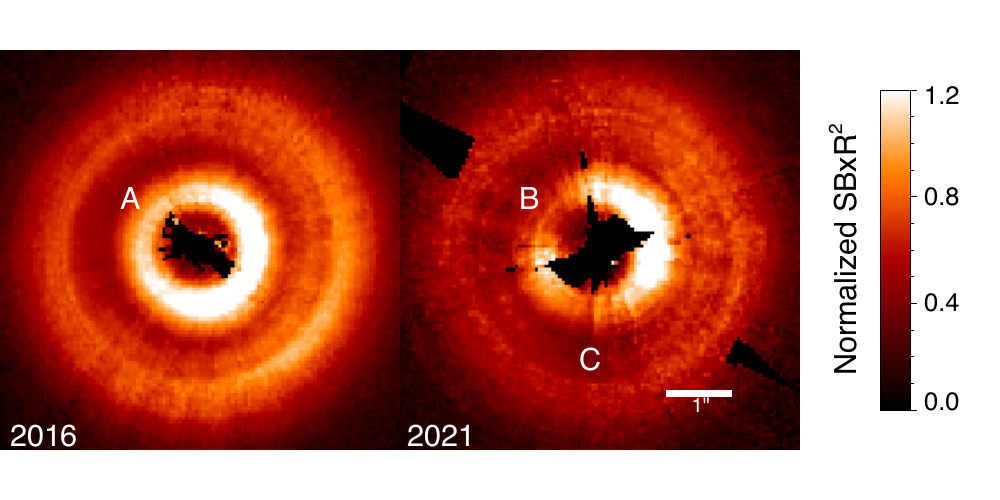}
    \caption{STIS visible light images of the TW Hya disk in 2016 and 2021. (top) Each panel is labeled with inidividual PSF subtracted visits of TW Hya taken in 2021 behind BAR5 and WEDGEA1.0 and multiplied by $R^2$ to highlight the rings and gaps present in the system. Diffraction spikes and areas blocked by the occulters are masked out. We recover the overall nature of the disk irrespective of the spacecraft orientation.  (bottom) Comparison of the disk between 2016 and 2021. In 2021, the smooth shadow features (seen at a PA of $\sim$50$\arcdeg$ in 2016 and marked as "A") have split into two shadow features marked by the "B" at 49~$\arcdeg$ and "C" at 175~$\arcdeg$.}
    \label{fig:f1b}
\end{figure}

\section{Observations} \label{sec:obs}

TW Hya was observed on 7 June 2021 with the STIS Instrument \citep{woodgate98}, using the CCD and 50CORON aperture as part of GO\#16228 (PI: Debes). The same PSF reference for TW Hya, HD 85512, was used as in \citet{Roberge:2005} and D17. TW Hya was observed for two consecutive orbits with spacecraft orientations separated by 17$\arcdeg$, followed by an orbit on HD 85512, and a final orbit on TW Hya separated by 32$\arcdeg$ from the first TW Hya orbit. Each star was first placed behind the BAR5 aperture location which has a half-width of 0\farcs15 with inner working angles of $\sim$0\farcs2 \citep{glennisr,debes19}. For TW Hya, 5$\times$110~s exposures were obtained per orbit behind BAR5, with sufficiently short exposure times chosen to avoid saturating the detector at the mask edge. The PSF reference was dithered perpendicular to BAR5 by $\pm$12.5~mas to account for the 0.25~pixel target acquisition non-repeatability \citep{glennisr}. We obtained 5$\times$11.2~s exposures at each dither location. Next, each star was placed behind the WEDGEA1.0 occulter to ensure sufficient signal-to-noise (S/N) in the outer disk without saturating the detector. The WEDGEA1.0 exposures were 3$\times$425~s per orbit with TW Hya and 22$\times$32~s exposures for HD~85512. The total integration time for TW~Hya was 1100~s behind BAR5 and 2550~s behind WEDGEA1.0. This procedure replicates the other STIS images obtained in 2015 and 2016 (D17). 

Unfortunately, a delay in guide star acquisition rendered the final orbit of TW Hya unusuable, since the star was not placed behind the proper apertures. The purpose of the multiple orbits of TW Hya are primarily to ensure nearly 360$^\circ$ coverage on the disk exterior to 0\farcs5. The result of the failed orbit is a modest decrease in disk S/N and a smaller angular coverage at small angular radii as seen in Figure \ref{fig:f1}. However, as we show below, this did not significantly impact our results when comparing to previous epochs at radii $>$40~au (0.67$\arcsec$).

We followed the PSF subtraction laid out in D17, which we briefly review. First, the stellar center was determined for both target and PSF reference. We then iteratively registered and scaled the PSF reference, minimizing subtraction residuals within a mask that included the diffraction spikes of the star and subtracted off the PSF reference. We note that the ratio of TW Hya's flux to HD~85512 across the STIS CCD bandpass was higher in 2021 between 0.2-1.0$\micron$~by 24-31\%. Compared to 2015 and 2016 where the best fit ratio of TW Hya's flux to HD~85512 was 0.052 and 0.055, we find that the best scaling between the two stars is now 0.0683. 

TW Hya has been known to be variable by $\sim$20-30\% in the visible on periods close to $\sim$3 days \citep{mekkaden98}. To investigate whether this magnitude of brightening might be expected in the STIS bandpass, we downloaded existing contemporaneous STIS/G430L and STIS/G750L spectra taken with the 52x2 slit of TW Hya available in MAST and estimated the count rate on the STIS detector for the 50CORON mode for each spectrum using the STIS ETC. The spectra came from four epochs (28 Jan 2010, 4 Feb 2010, 28 May 2010, and 18 Apr 2015) as part of programs 11608 (PI: Calvet) and 13775 (PI: Espaillat). The varying flux in the blue continuum and in emission line intensity of the different spectra account for peak-to-peak changes in predicted count rates on the STIS detector across the four epochs of 22\%, in line with the variability observed. Additionally, we inspected the acquisition images for TW Hya (which used the F28X50LP filter) between 2016 and 2021. Even though the filter throughput for the acquisition images is for flux beyond 5000~\AA, aperture photometry of the acquisition images showed a 22\% increase in TW Hya's brightness between the two epochs. The change in SED for TW~Hya adds modest variability in how well the target PSF reference matches in color to TW Hya, which could result in more or less PSF subtraction residuals.

With the scalings, we recover the disk again, seen in Figure \ref{fig:f1}. We show reductions of each individual visit to demonstrate the impact of PSF subtraction residuals. Residual features are present only in the detector frame of the observations and will appear to rotate when in a sky oriented frame. While some modest features appear to be due to PSF residuals, they are at the level of our estimated noise per pixel and do not impact our measurements signficantly for the analysis described in \S \ref{sec:az}.

As noted in D17 and \citet[][hereafter VB17]{boekel17}, when one removes the 1/R$^2$ dependence of disk illumination from the star, a gap is present at $\sim$80~au, with an inner ring peaking at $\sim$30~au, with another gap interior to that as shown in the 2016 image in Figure \ref{fig:f1}. There  is an azimuthal surface brightness asymmetry, which we interpret as due to a shadow (D17). In past epochs, this shadow asymmetry was well described by a smooth sinusoidal curve. The position angles (PA) of the peak and trough of the asymmetry have, to date, moved with constant angular velocity across several wavelengths in the visible and NIR. Using the 15.9~yr predicted period from D17 and setting $t_{\mathrm{o}}$ to equal the first observation of the TW Hya disk, we find that roughly half of the predicted orbit for the shadow has not been covered, and the June 2021 observations mark a similar phase of the orbit to archival NICMOS polarimetric observations \citep{poteet18} and archival medium band NICMOS total intensity observations of the disk (D17) taken in 2004 and 2005, both of which are of lower S/N compared to the observations with STIS.

We created an azimuthal average surface brightness profile of the 2021 dataset, converted it into physical flux units per pixel, and compared the resulting fluxes with azimuthal averages of the 2016 observations of the TW Hya disk. Due to the broad bandpass of the 50CORON mode, we estimate the conversion from counts s$^{-1}$ on the detector to surface brightness by using the stsynphot package and the average spectrum of TW~Hya obtained with STIS. We find that the conversion is 1 mJy arcsec$^{-2}$=3.54 DN/s. We find that the flux profile of TW~Hya is slightly different in 2021 compared to 2016, primarily fainter by 20\% interior to 1$\arcsec$\ as well as beyond 1.5$\arcsec$.

\section{Azimuthal Surface Brightness Evolution} \label{sec:az}
Unlike the observations taken in 2016, the inner ring shows two narrow dark lanes at PA of $\sim$50$^\circ$ and $\sim$180$^\circ$. The peak of the azimuthal asymmetry has moved counter-clockwise toward due North as predicted by the shadow rotation model presented in D17. To quantify these changes, we replicate the  methods of D17 by measuring the azimuthal surface brightness (ASB) of the disk as a function of PA and radius for the 2021 epoch and compare those to the previously published profiles in 2016.

To obtain azimuthal asymmetries, we divided each ASB profile for a given radius by the mean SB at that radius to allow for unbiased radial averaging.  We chose angular radii that matched the measurements made in D17, and converted to physical differences with the new Gaia EDR3 parallax \citep{gaia1,edr3}: seven radial locations centered at 0.46\arcsec\ (28~au), 0.66\arcsec\ (40~au), 0.89\arcsec\ (54~au), 1.14\arcsec\ (68~au), 1.47\arcsec\ (88~au), 1.83\arcsec\ (110~au), and 2.36\arcsec\ (142~au), with widths of 0.2\arcsec\ (12~au), 0.2\arcsec\ (12~au), 0.25\arcsec\ (15~au), 0.25\arcsec\ (15~au), 0.41\arcsec\ (25~au), 0.30\arcsec\ (18~au), and 0.56\arcsec\ (34~au) respectively.  The locations are within and near the two gaps seen in scattered light at 22 and 88~au with sufficient S/N.  We estimated the root-mean square uncertainty per azimuthal bin by calculating both the average uncertainty per pixel as estimated by the total counts within each pixel or by the standard deviation of counts within an azimuthal bin, whichever was larger. At these radii, the profiles in 2016 beyond 40~au have a median S/N of 19 per azimuthal bin, while for 2021 the median S/N per azimuthal bin is 18.

\begin{figure}
    \centering
    \plotone{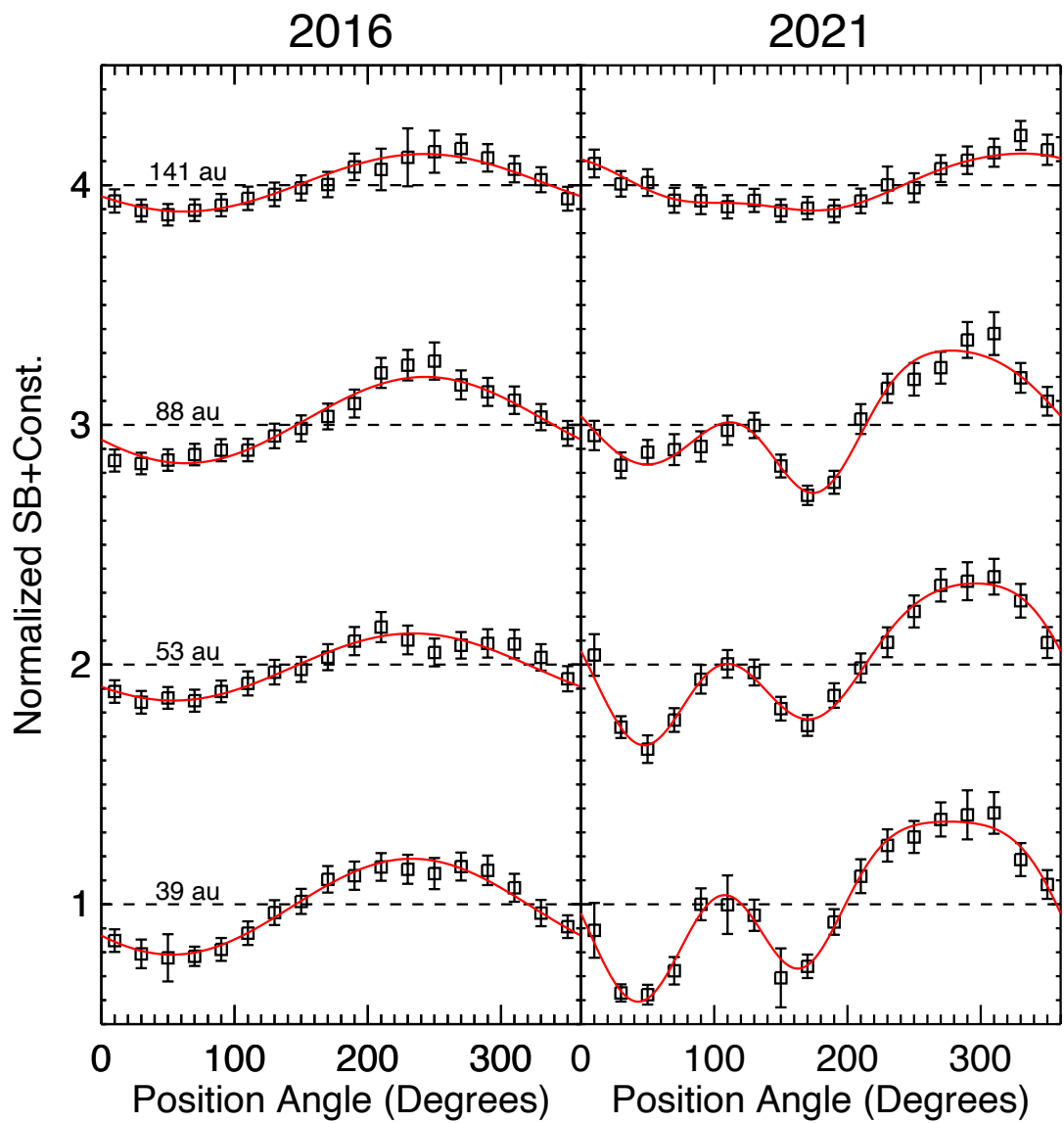}
    \caption{Comparison of the 2016 and 2021 epochs of TW Hya's azimuthal surface brightness distribution as a function of radius. On the left, the 2016 epoch shows that the azimuthal brightness is well defined by a cosine or sine curve with a clear peak and trough position. The 2021 epoch to the right shows a departure from this behavior. The presence of two quasi-Gaussian shadows are present. The red lines show cosine fits to the data taken in 2016 from D17 and two-Gaussian fits to the data taken in 2021.}
    \label{fig:azimuth1}
\end{figure}

Figure \ref{fig:azimuth1} shows the results for the 2016 and 2021 epochs. The simple cosine fit used in D17 is no longer sufficient to explain the observed ASB profile in 2021, thanks to the apparent presence of two shadow features. Figure \ref{fig:azimuth1} shows that the two shadows diminish in depth at larger radii, such that they are no longer distinguishable as two components and that the peak surface brightness increases in PA with radius. We thus fit a cosine at 141~au and the PA of the peak surface brightness is 330$^\circ\pm10^\circ$, in tension ($\sim2.3\sigma$) with the 2005 measurement from the NICMOS medium band data at a similar phase of the predicted motion reported in D17. We note that the 2004 NICMOS total intensity data derived from the POL filters in D17 and P19 showed a similar lower than expected PA for the given period.

There are three possibilities that can explain the new observations. First, there are new features caused by the recent appearance of additional shadowing structures unrelated to what has been seen before, such as new dust clumps. Secondly, warps due to companions that cast shadows can appear when the planet is on the near side of the disk from the perspective of the observer \citep{nealon19}. This is possible since we are seeing the phenomena at a PA that is close to the major axis of the disk at 152$\arcdeg$ as measured by CO data \citep{huang18}. Thirdly, it is possible that these two shadows arise from two mutually inclined disks that have precession timescales which are slightly different and casting overlapping shadows, thus mimicing a single shadow over the previously observed epochs.

The ASB profiles can be fit by combinations of shadows and cosine functions. At present it is difficult to find a unique solution between these possibilities. We therefore focus on the simplest case--that there are two shadows that previously overlapped but are moving at different angular velocities and partially separated in 2021. Thus, we can approximate the shape of the shadows as a linear combination of Gaussian curves.

In order to fit the two shadow features simultaneously, we use CURVEFIT.pro, an IDL nonlinear least squares fitting routine assuming two Gaussian shadows, with constant offset ($C$), shadow depth ($D_{\mathrm i}$), shadow PA ($\theta_{\mathrm o,i}$), and shadow angular width ($\sigma_{\mathrm i}$):

\begin{equation}
    F(\theta)=C+\sum_{i=1}^{2}{}D_{\mathrm i}\exp{\frac{-(\theta-\theta_{\mathrm o,i})^2}{2\sigma_{\mathrm i}^2}}
\end{equation}
where we convert the shadow width to an angular full-width at half maximum $FWHM_{\mathrm i}\approx2.3548\sigma_{\mathrm i}$. The fits are shown in Figure \ref{fig:azimuth1} and show no significant residuals.

\begin{figure}
    \centering
    \plotone{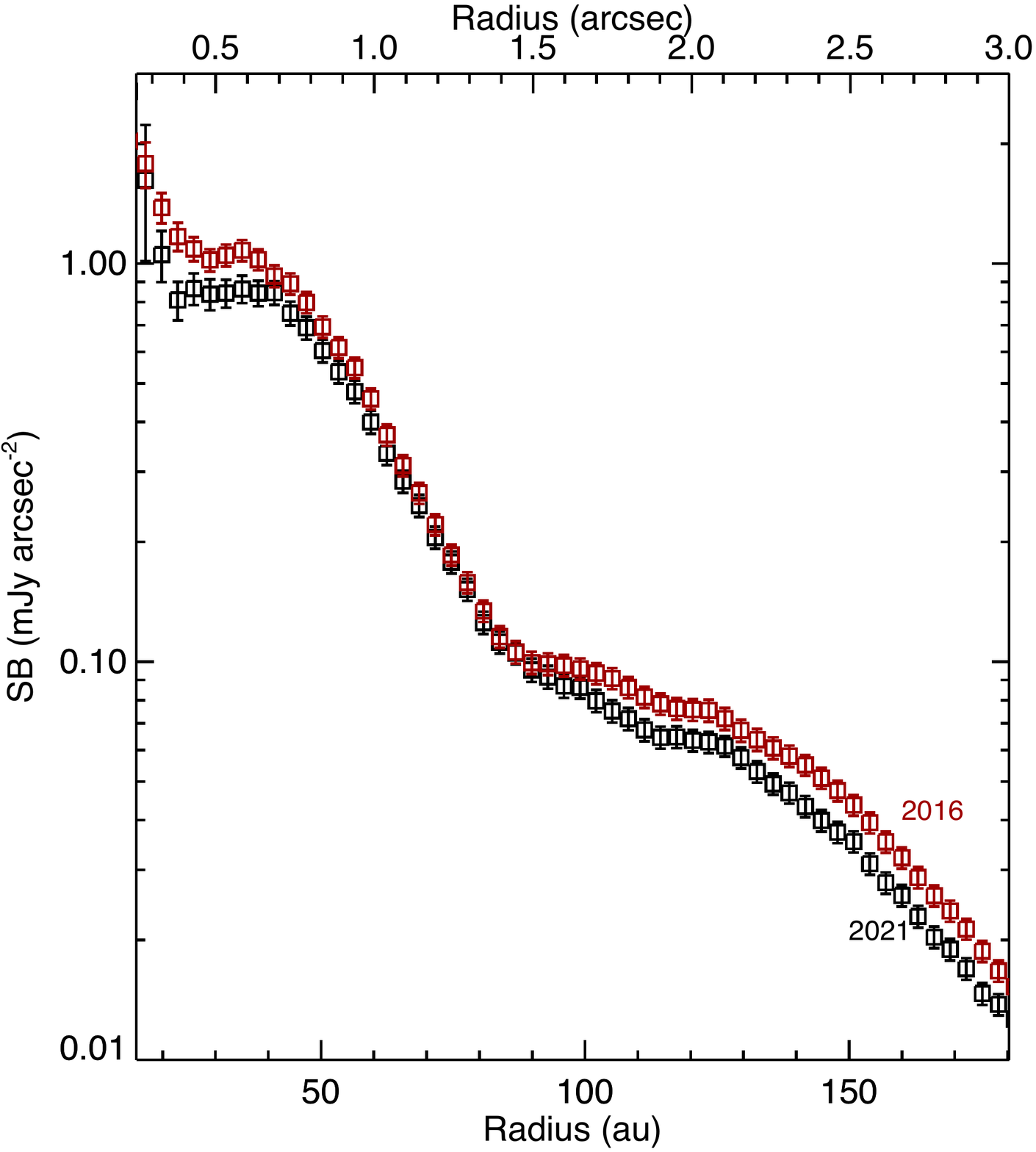}
    \caption{Comparison of the azimuthally averaged surface brightness profiles of TW Hya from 2016 and 2021.  At small radii and larger radii, the surface brightness has slight differences of 20\%}
    \label{fig:radial}
\end{figure}

\begin{figure}
    \centering
    \epsscale{0.8}
    \plotone{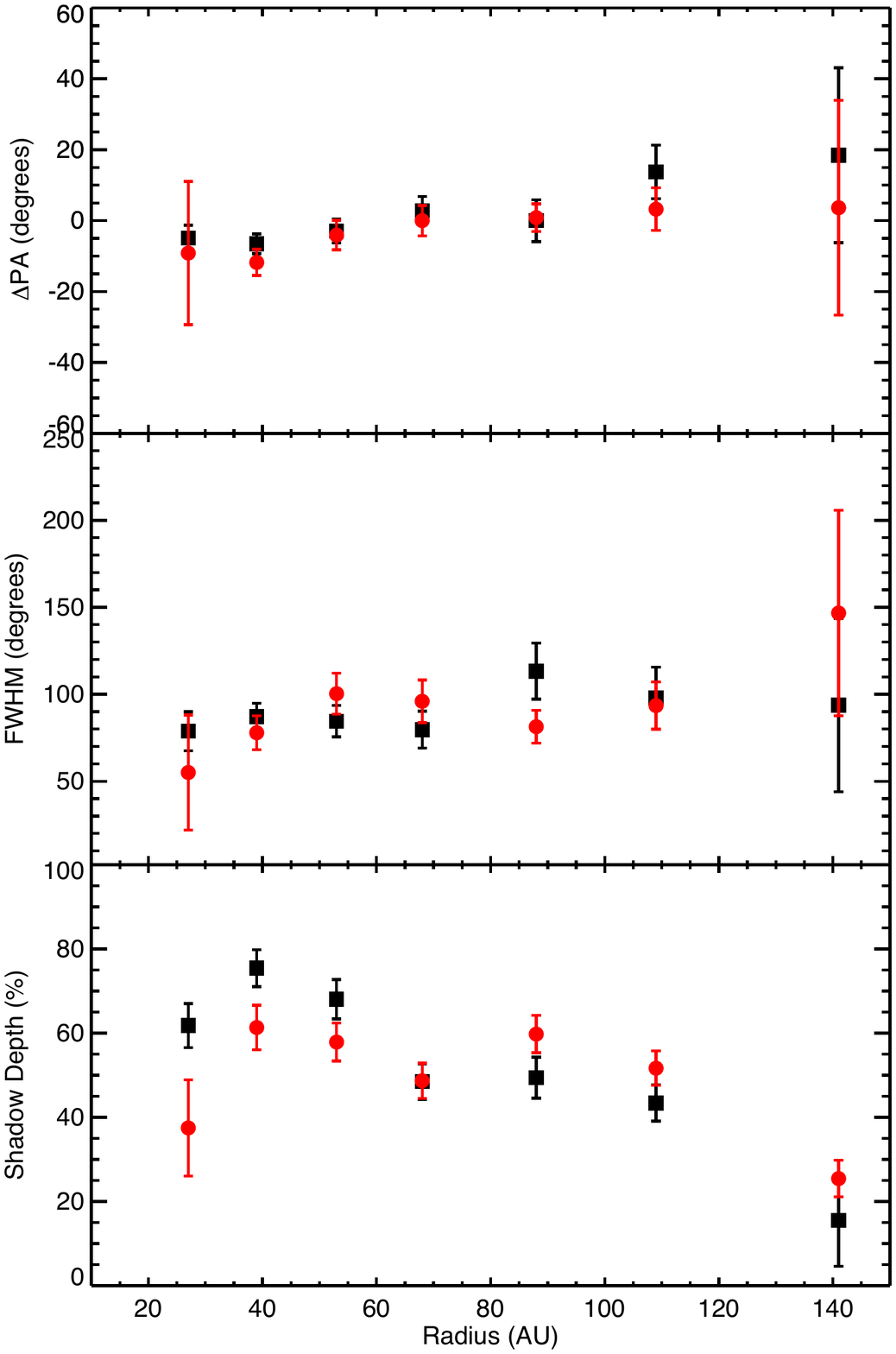}
    \caption{Fit parameters for shadow A and B as a function of radius. Shadow A is represented by black squares while shadow B is represented by red squares. (top) Relative PA from the median. (middle) The measured FWHMs of the two shadows do not show significant changes with radius. Note that at the largest radius, the uncertainty in FWHM is large since shadow position and width are degenerate when the shadows overlap. The shadows decrease in depth as a function of radius.}
    \label{azimuthdetail}
\end{figure}

Figure \ref{azimuthdetail} shows the PA, widths, and depths of the shadows as a function of radius. Shadow A has a median PA of 49$\arcdeg$, while Shadow B has a median PA of 175$\arcdeg$. The shadows show a slight increase in PA in the counterclockwise direction as radius increases, which may be indicative of warping in the disk surface.

The FWHM of the shadows is broad: the median FWHM of shadow A is 87$\arcdeg$, shadow B is 93$\arcdeg$. These widths can explain why the shadow might have appeared to be singular previously if the separation of the two shadows was less than their FWHMs. Finally, the depth of the shadow changes with radius, starting at depths of near 80\% and decreasing to 20\% by 140~au.

\begin{figure}
 \centering
    \epsscale{0.8}
    \plotone{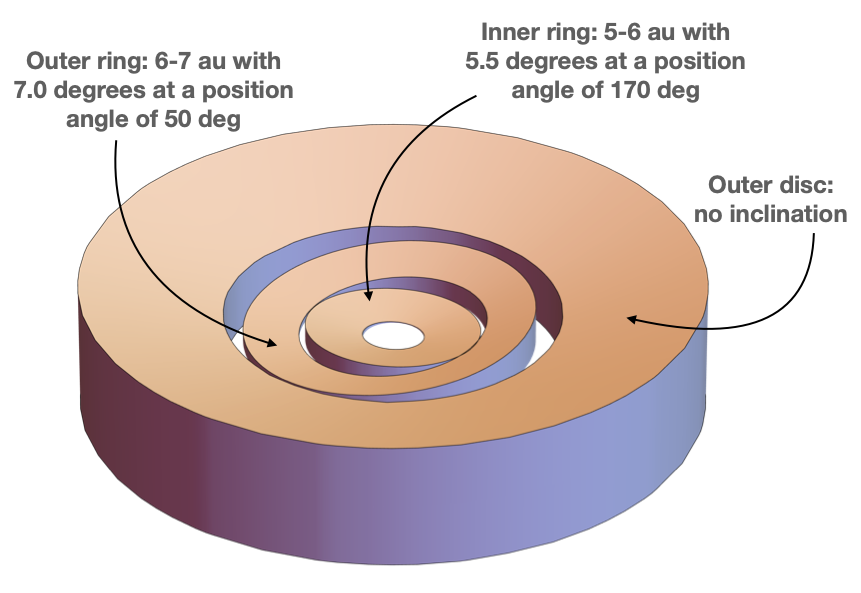}
    \caption{Schematic of our proposed model to explain the evolution of the TW Hya shadow. We consider two mutually inclined rings between 5-6~au and 6-7~au with slightly different inclinations relative to the outer disk as an origin to the shadows we observe in 2021.}
    \label{fig:schematic}
\end{figure}


\begin{figure}
    \centering
    \epsscale{0.8}
    \plotone{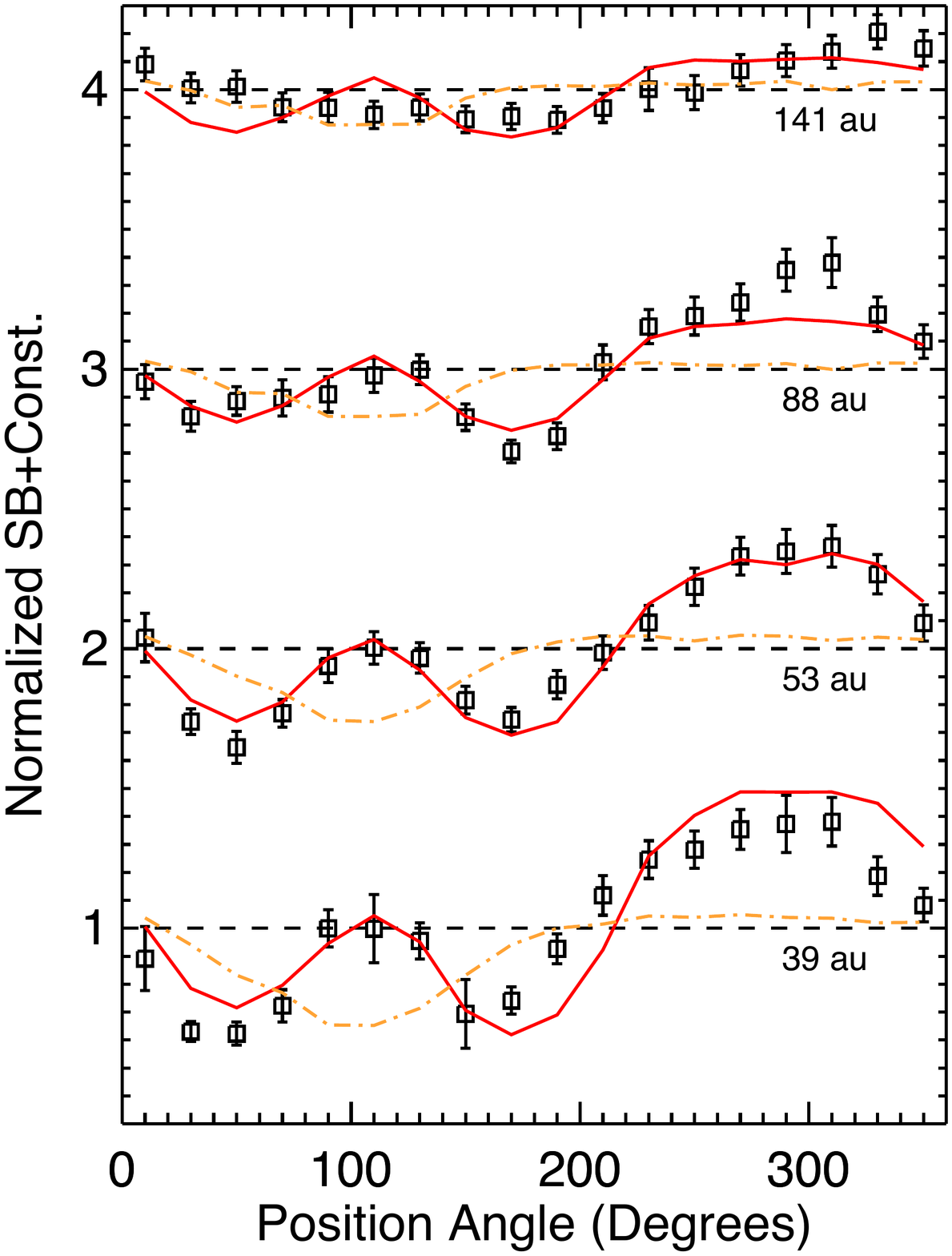}
    \caption{Comparison between azimuthal profiles at different radii taken in 2021 and different radiative transfer models of shadows from inclined rings. The black squares are the data with error bars, while the orange dashed line is the profile from a single inclined ring casting a shadow. The red solid line is the profile derived from two inclined rings but with differing PAs. The two-ring model better reproduces the observed profiles than a single ring model.}
    \label{fig:modeling}
\end{figure}

\section{Implications for the Origin of the TW Hya Shadows} \label{sec:implications}
We review the following characteristics and evolution of the disk shadow in order to provide a summary of the current observational constraints. We will restrict ourselves to those behaviors that are seen in the STIS data, since this provides the set of constraints that cannot be explained by changes in behavior of the shadow with wavelength, instrumental setup/systematics, or disk polarization.

From 2000-2016 the shadow covered 180$\arcdeg$ of azimuth and the motion of both edges of the shadow were consistent with constant angular motion. The shadow depth was constant over this time period and with radius. The PA of the shadow at r$<$50~au did not always match that of outer parts of the disk. The disk shadow detected prior to 2016 may have evolved into two shadows by 2021. The behavior of the shadow between 2000, 2015, and 2016 suggested that the shadow rotated counter-clockwise with a period of $\sim$15.9 years. The behavior of the shadow in 2021 may mean that either rotation or oscillatory behavior of the shadow is possible, but the overall period still seems to be $\sim$15~yr, limited somewhat by sparse time sampling of the shadow's motion with STIS-only epochs. In 2021, there is a trend where the depth of the shadows changes quasi-monotonically with radius, with the shadow becoming less pronounced further out in the disk. This behavior is not as pronounced in other STIS epochs.

In order to put a quasi-physical explanation to this new shadow behavior, we begin in \ref{subsec:fiducial} to model the disk without any shadowing structures as a baseline disk model to compare to two different scenarios, a single inclined ring casting a shadow and two mutually inclined rings casting shadows. In \ref{subsec:onering} we attempt to fit the data with one inclined ring. This is because a tilted ring is physically motivated (e.g. D17), conserves angular momentum, in some cases has been observed \citep{marino15} and has been inferred in other disks \citep{muro-arena20}. We then use the rapid radiative transfer code MCFOST \citep{pinte06,pinte09} to model the shadow cast by an inclined ring to try and change the location, shape and inclination of this ring to match the flux profiles in Figure \ref{fig:azimuth1}. Ultimately, we show the best fitting inclined inner ring cannot fit the observed ASB and investigate the impact of adding a second inclined ring in \ref{subsec:tworing}. There we find a good match to both the SED of the disk and the visible scattered light images, which suggests that this scenario is the most likely explanation for the behavior we observe.

\begin{figure}
    \centering
    \epsscale{0.8}
    \plotone{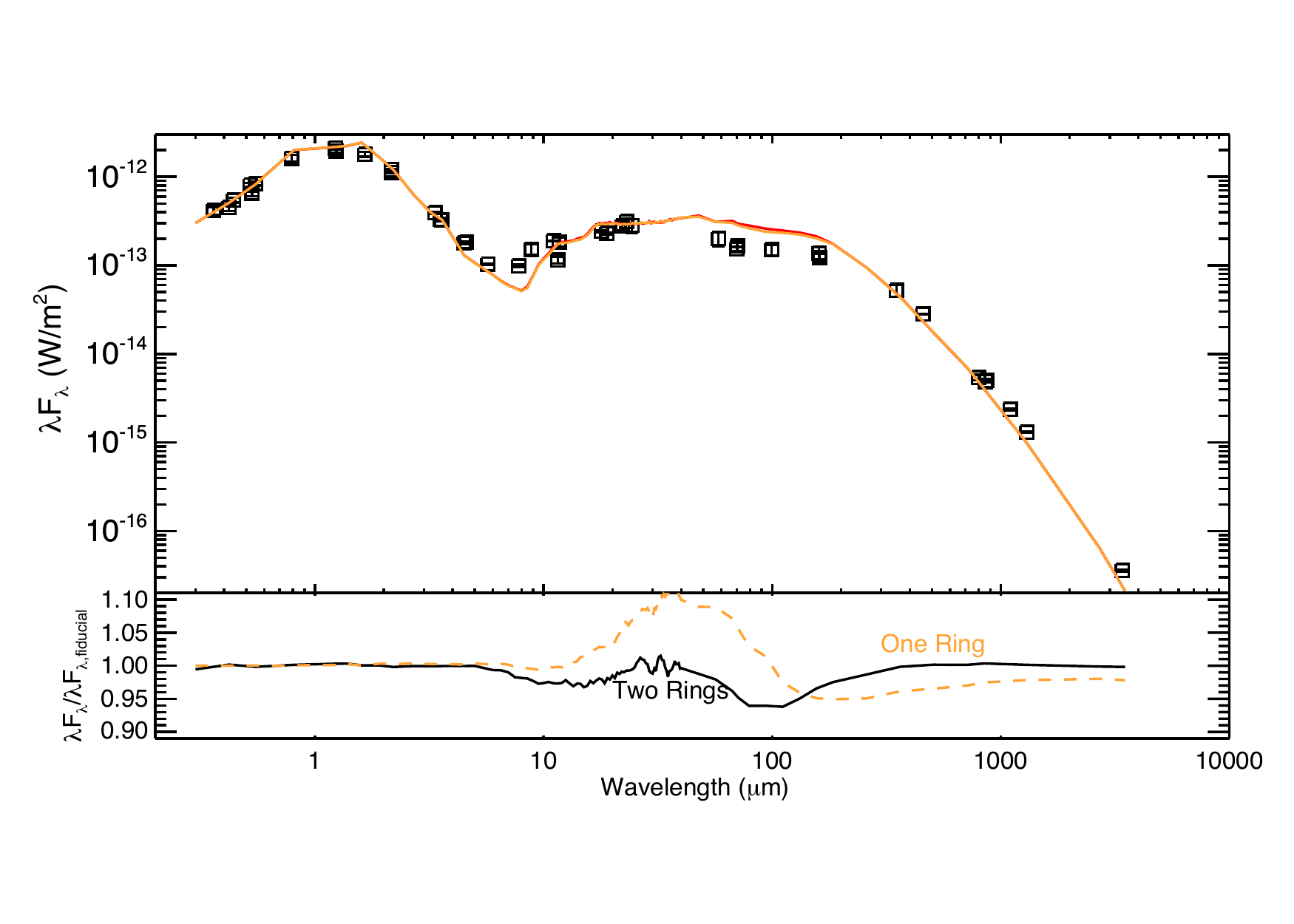}
    \caption{(top) Black points represent the SED of TW Hya. The red curve is the model SED from our fiducial radiative transfer model. (bottom) Comparisons between the fiducial model SED and those of the single and double ring models.  All models are similar to within 5-10\% of the fiducial model.}
    \label{fig:sed}
\end{figure}

\subsection{A fiducial Model for the TW Hya disk}
\label{subsec:fiducial}
The first step is to create a disk model that is broadly consistent with both the visible total intensity scattered light morphology and the spectral energy distribution measured for TW Hya without any shadows. 
To that end, we recreated the model proposed by VB17 with MCFOST. VB17 was focused on simultaneously fitting the polarized intensity radial surface brightness profile of the disk near-IR wavelengths with the TW Hya SED. This model was also derived from work done by \citet{menu}, which was a model focused on self-consistently fitting multi-wavelength interferometry data with the TW Hya SED. Both models relied on a 1-D axisymmetric disk structure with a vertical scale height dictated by hydrostatic equilibrium. 
Both models assumed that the inner gas disk was truncated with a rounded inner taper and that the largest dust grains were decoupled from the gas radial density profile. VB17 additionally added an outer gas density taper at a larger radius than \citet{menu}, used the ALMA 870~\micron\ radial surface brightness profile \citep{andrews16} as a proxy for the large dust radial profile, and added gaps in the form of density decrements at particular radii to reproduce the polarized intensity radial scattered light surface brightness profiles observed with SPHERE. 
For the purposes of our models, we assumed that the outer disk has an inclination of 0${\arcdeg}$ and a position angle on the sky of 155$^{\arcdeg}$. While the true inclination of the outer disk is likely closer to 5$\arcdeg$, the impact to both the scattered light images and SED are negligible.

While we consider a very similar disk structure, we make some alternate choices that appear to retain the basic observational features of the TW Hya disk, while making the model more amenable to calculating the radiative transfer of tilted inner rings. We are not trying to make a disk model that matches all available observations but a plausible model that can be used to investigate the impact of tilted inner rings on the observed surface brightness. We leave a deeper investigation into the degeneracies and constraints imposed by disk shadows for later work. In general, we find decent agreement with those of VB17, though we point out differences where they occur.

\subsubsection{Radial Surface Density of Gas and Small Dust}
The previous two models of the TW Hya disk calculated an analytical radial surface density profile for the gas while assuming a gas to dust ratio for the small dust grains and calculated a self-consistent scale height for the disk as a function of radius. Our approach is to calculate a fully analytical gas density distribution with the following form:
\begin{equation}
\rho(r,z)= \Gamma_{in}(r)\frac{\Sigma(r)}{\sqrt{2\pi}H(r)}f(r)\Gamma_{out}\int_{-\infty}^{\infty}\exp{\left(-\frac{z^2}{2H(r)^2}\right)}dz
\end{equation}
where $\Gamma_{in}(r)$ is the interior density taper, $\Sigma(r)$ is the surface density profile, $f(r)$ is a multiplicative combination of density gaps to mimic the gaps observed in various scattered light datasets, $\Gamma_{out}$ is the exterior density taper, and $H(r)$ is the vertical scale height of the gas (VB17, and references therein).

The interior taper replicates the format in VB17 and is:
\begin{equation}
\Gamma_{in}(r<r_{in})=\exp{\left[\left(\frac{1-r/r_{in}}{w}\right)^3\right]}.
\end{equation}

The surface density profile is a power-law:

\begin{equation}
\Sigma(r)=\left(\frac{r}{r_{in}}\right)^{-p}
\end{equation}
where we choose p=-0.7 and $r_{in}$ is the location where the interior taper takes effect.

The scattered light profile of TW Hya clearly shows the evidence of gaps or other perturbations in the upper layers of the disk, some of which correspond to features also seen in the sub-mm where the disk is optically thin beyond a few 10s of au\citep[][VB17]{andrews16,debes13}. We reproduce these features by including Gaussian density perturbations following VB17:

\begin{eqnarray}
f_{i} = & 1-d_{i}\exp{\left[-\frac{(r-r_{c,i})^2}{2\sigma_{in,i}^2}\right]} & \mathrm{for}\ r < r_{c,i} \\
f_{i}  = & 1-d_{i}\exp{\left[-\frac{(r-r_{c,i})^2}{2\sigma_{out,i}^2}\right]} & \mathrm{for}\ r > r_{c,i} \\
\end{eqnarray}

While we assume the perturbations are in density, we do not include similar perturbations in scale height which might also occur in a system undergoing gap formation \citep{bi21}.

We include a taper to the outer part of the gas density to match the sharp drop in surface brightness seen in scattered light and in deep observations of the outer gas disk in the sub-mm \citep[D17,][]{ilee22}. We assume the following functional form for the taper that occurs at $r>r_{out}$:

\begin{equation}
\Gamma_{out} = \exp{\left[\frac{-(r-r_{out})^2}{2\sigma_{out}}\right]} \mathrm{for}\ r > r_{out} \\
\end{equation}

Finally, we assume an analytical scale height such that $\frac{H}{r}$=0.05 at 0.5~au and that the disk has a flared geometry:
\begin{equation}
H(r)=0.05 \left(\frac{r}{0.4 au}\right)^{1+q}
\end{equation}
where $q=0.15$, shallower than the flaring expected for a purely isothermal disk \citep{chiang97}. We choose this analytical form to best match the SED and the surface brightness of the observed disk. To realize the density distribution we create a cylindrical grid in the density with 4.8$\times10^{5}$ cells logarithmically spaced in radius. 

\subsubsection{Stellar Parameters}
The fundamental stellar parameters of TW~Hya cover a wide range in mass and T$_{\mathrm eff}$, primarily because of the intrinsic difficulty of determining the parameters of young stars. For the purposes of the fiducial model, we assume similar values to those assumed in VB17, accounting for the larger distance to TW~Hya determined by Gaia DR3\citep{edr3}. In MCFOST, we chose T$_{eff}$=4000~K and $\log g$=4.0, assuming a mass of 0.6~\Msun. These parameters are quite close to those determined by recent studies of TW Hya's temperature and gravity \citep{sokal18}. They are also consistent with the inferred M$_{\star} sin~i$ from CO measurements of Keplerian velocities \citep{huang18,teague18}.

\begin{deluxetable}{ccccc}
\tablecaption{Table of parameters used to describe the density distribution of gas in the TW Hya disk}
\tablehead{\colhead{Parameter} & \colhead{VB17} & \colhead{Fiducial} & \colhead{One Ring} & \colhead{Two Rings} }
\startdata
p  & -0.75  &   -0.70   & -0.70  &  -0.70 \\
w  & 0.45~au   & 0.50~au   &   0.50~au  &   0.50~au \\
r$_{in}$ & 2.7~au &  3~au   &     3~au     &    3~au \\
q  & ...\tablenotemark{a} & 0.15  & 0.15 &   0.15 \\
r$_{out}$ & 104~au  & 100~au &    100~au   &   100~au \\
$\sigma_{out}$ & 50~au  & 50~au  &    50~au   &    50~au \\
\enddata 
\tablenotetext{a}{q was not a free parameter in VB17's model but was calculated via hydrostatic equilibrium}
\end{deluxetable}

\begin{deluxetable}{ccccc}
\tablecaption{ \label{tab:parameters} Table of parameters used to describe the density gaps in the TW Hya disk.}
\tablehead{\colhead{Parameter} & \colhead{VB17\tablenotemark{a}} & \colhead{Fiducial} & \colhead{One Ring} & \colhead{Two Rings}
}
\startdata
d$_1$ &   0.44& 0.44 & 0.44 & 0.42 \\
r$_{c,1}$ & 94~au & 85~au & 85~au & 85~au \\
$\sigma_{in,1}$ & 16.7~au & 12~au & 12~au & 10~au \\
$\sigma_{out,1}$ & 17.8~au & 30~au & 30~au & 30~au \\
d$_2$ & 0.61 & 0.65 & 0.65 & 0.9 \\
r$_{c,2}$ & 23~au & 24~au & 24~au & 24~au \\
$\sigma_{in,2}$ & 4.2~au & 8~au & 8~au & 10~au \\
$\sigma_{out,2}$ & 20~au & 8~au & 8~au & 8~au \\
d$_3$ & 0.81 & 0.81 & 0.81 & 0.92 \\
r$_{c,3}$ & 6.7~au & 6.7~au & 6.7~au & 6.7~au \\
$\sigma_{in,3}$ & 3.3~au & 3.3~au & 3.3~au & 3.3~au \\
$\sigma_{out,3}$ & 14.4~au & 18.7~au & 18.7~au & 18~au \\
\enddata 
\tablenotetext{a}{corrected for Gaia DR3 distance}
\end{deluxetable}

\subsubsection{UV Excess}
For completeness, we include a UV excess to TW Hya that fits the visible observed photometry. In reality, TW~Hya's UV flux can vary by factors of two or more. We verified that the addition or subtraction of the UV excess in MCFOST has a minimal impact on the predicted scattered light surface brightness and the SED.

\begin{figure}
    \centering
    \epsscale{0.8}
    \plotone{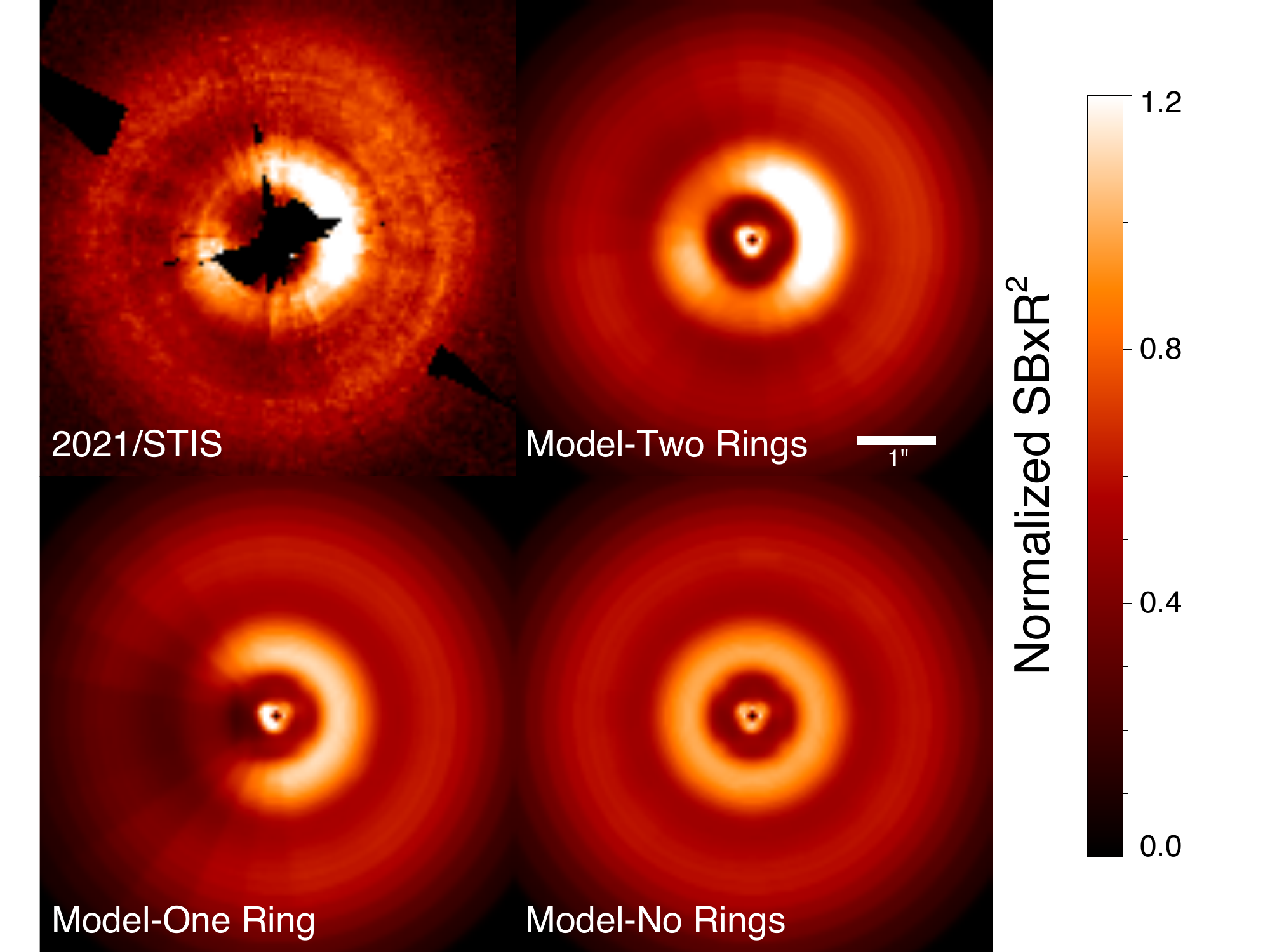}
    \caption{Comparison between the 2021 STIS image of the TW Hya disk (upper left) compared to different models, including a fiducial model with no tilted inner rings, a singular tilted inner ring, and two tilted inner rings.} 
    \label{fig:imgmodeling}
\end{figure}

\subsubsection{Grain Properties}
We initially replicated the composition of dust used in VB17, namely using a mixture of 80\% amorphous magnesium iron olivine-style silicates (MgFeSiO$_4$) \citep{dorschner95} and 20\% amorphous carbon \citep{rouleau91}. We note that VB17 made use of different amorphous carbon optical constants but it does not appear to greatly impact the results \citep{priebisch93}. We selected an MRN distribution for the dust, using a minimum grain size of 0.1~$\micron$\ and maximum grain size 0f 10000~$\micron$. This effectively assumes that the large grains trace the smaller grains, which is not supported either by interferometry in the mid-IR or in the sub-mm, nor follows the models of VB17 or \citet{menu}, where large grains followed a separate density pattern. That said, since our model focused solely on reproducing a 1-D SED and 2-D scattered light model, we assume that this choice has only a minor bearing on our results with respect to reproducing shadows. We assume that the gas to dust ratio in the disk is 100. The composition of dust considered by VB17 results in a surface brightness that is too faint to match the STIS observations. This can be ameliorated by increasing the scale height of the disk, but resulted in a poorer fit to the SED. After some trial and error we use a composition more similar to that of \citet{debes13}, which is based on SED fitting by \citet{calvet}. For our fiducial model, we assume a mixture of 60\% Olivine and 40\% water ice, but we note that this is likely a non-unique solution.

\subsubsection{MCFOST parameters}
We input the analytical 3-D density profile as a grid into MCFOST and then selected 10$^{7}$ for the number of photon packages used for the scattered light images. Additionally, we selected a pixel size equivalent to the platescale of a STIS pixel, we fixed the disk mass at 5$\times$10$^{-4}$\Msun, the inner radius at 0.38~au and the outer radius 222~au to match those of VB17 accounting for the updated DR3 Gaia distance. We assumed Mie scattering for our calculations and a single wavelength at 7000~\AA\ that mimics the expected central wavelength of the observations. The STIS CCD bandpass extends from 2000-10000~\AA\ and thus the central wavelength of an observation is dependent on the underlying SED of the source, which to zeroth order is equivalent to the SED of TW~Hya, since the dust scatters light nearly constantly across visible wavelengths \citep{Roberge:2005}. Additonally, we calculate a model PSF from the TinyTim software \footnote{https://www.stsci.edu/hst/instrumentation/focus-and-pointing/focus/tiny-tim-hst-psf-modeling}, assuming K7 stellar spectral type and convolve it with the model images to compare with the STIS data.

\subsubsection{Fiducial Model fidelity}
Figure \ref{fig:sed} shows a comparison of the fiducial model SED against the observed photometry reported in \citet{menu}. The model SEDs for the fiducial model and the subsequent tilted ring models we consider are virtually identical, demonstrating that such a structure has a relatively minimal impact on the one dimensional unresolved SED. Our fiducial SED fits about as well as those of \citet{menu} and VB17, though we did not attempt to optimize the fit--the reduced $\chi^2$ for the fiducial fit is 3.22. Overall, the SED is close to that observed, with an overprediction of flux between 50-200~\micron, and slight underprediction of flux in the sub-mm regime. Additionally, Figure \ref{fig:imgmodeling} shows a comparison between the observed STIS image in 2021 and the fiducial model. The agreement between both the SED and the general surface brightness of the disk is excellent. Since the fiducial disk is face-on and with no shadows, the ASB is constant.

\subsection{Shadows from Single Tilted Rings}
\label{subsec:onering}
We next consider the resulting shadows caused by a single tilted ring at small inclination. To tilt a restricted section of the disc, we incline the midplane as required and ensure we are measuring the perpendicular distance from the midplane to determine the density at a given location. If the inclination of the ring is too small or the flaring of the disk large the shadow will have a small depth and not extend over a large range of radii in the disk. As inclination increases for fixed flaring, the shadow deepens and widens. At moderate inclination the single shadow breaks into two as the star illuminates the outer disk in between the line of nodes for the inner disk and approaches the situation considered by \citet{min17}. However, in these cases the depth of the remaining shadows is much deeper than observed for TW~Hya and does not match the detailed ASB. For that reason we consider a small inclination to the inner ring as the preferred scenario.  For our fiducial model, we find a ring at a position angle of 100$\arcdeg$ at a radius of between 5 and 6~au with an inclination of 7$\arcdeg$ that roughly replicates the depth, but not the detailed shape of the June 2021 ASB in Figure \ref{fig:modeling}. Further, we note that the width of the single disk shadow feature is narrower than what was observed previously for TW~Hya, implying that in past epochs the shadow feature seen was already indicating the presence of more than one tilted disk.
 
\subsection{Shadows from Two Tilted Rings}
\label{subsec:tworing}
We repeat this process but consider two independently oriented tilted rings (See Figure \ref{fig:schematic}). Here we choose two rings misaligned by 5.5\arcdeg\ and 7\arcdeg\ with respect to the outer disk plane, the first with a PA=170$\arcdeg$ extending from 5-6~au and the second from 6-7~au with PA=50$\arcdeg$. These distances are consistent with the $\sim$16~yr timescale of the existing shadow motions assuming the motions are Keplerian and the magnitude of the inclination is chosen to roughly match the depth of the observed shadows, but other configurations are likely also possible. Investigating the degneracies in the ring location and width is beyond the scope of this paper. If we place the rings too far out $r_{c}>15~au$, we cannot easily reproduce the surface brightness beyond 30~au.

We find that the inclusion of a second ring forces slight alterations to the fiducial inner gap depths to retain a good fit to the azimuthally averaged surface brightness profile. In particular, the 24~au gap needs $d=0.9$ as opposed to $d=0.65$ in the fiducial profile and the 6.68~au gap needs $d=0.92$ as opposed to $d=0.81$. 

With the inclusion of the second ring, the resulting model image matches the June 2021 image as well as the ASB (See Figures \ref{fig:imgmodeling} and \ref{fig:modeling}). We therefore demonstrate that two mutually inclined rings at different position angles well represent the observed disk images and represent the best explanation for the change in behavior. At earlier times, the PAs of the two rings were closer together and their resultant shadows overlapped, causing an apparent single shadow.

\subsection{Observational implications for two tilted rings}
The implied radial distances of $\sim$5-7~au for the two ring model are accessible via interferometry at visible and NIR wavelengths as well as with direct imaging. Such observations can provide a potentially useful constraint on the inner shadowing structure at wavelengths close to the STIS bandpass with our images having an effective wavelength close to the Johnson-Cousins $R$ band, but such constraints are heterogeneous in the literature. K-band GRAVITY observations of the inner disk edge (R$\sim$0.04~AU) are consistent with inclinations of $<20$\arcdeg \citep{gravity2021} a multi-wavelength treatment of interferometry spanning NIR to radio wavelengths suggested a disk structure with an inner radius of $\sim$0.8~au (corrected for the updated distance to TW~Hya) and a puffed up disk rim located at $\sim$3.6~au \citep{menu}. 

Polarized intensity imaging with SPHERE/ZIMPOL showed gaps/rings which could be coincident with the shadowing structures we consider, as did ALMA \citep[][VB17]{andrews16}. Interior to 28~au, the ZIMPOL observations had a tentative detecton of an inner disk between 2-6~au and a ring at $\sim$16~au. ALMA showed emission maxima at $\sim$3, 11, and 16~au with some additional substructure in a plateau around 29-38~au. Most of the locations interior to 15~au are consistent with the shadows, although more stringent constraints can be placed if the exact motion of the shadows is determined. 

Additional monitoring of the shadows' motion is needed to understand whether the 16~yr periodicity seen is still relevant and whether it is due to orbital motion, or mutually interacting misaligned rings. Any scenario that involves precession will require fairly massive planetary or substellar companions and short orbital periods to show periodicity on 16~yr timescales \citep{nealon20}.

\section{Conclusions}
\label{sec:conclusions}
We report new images of the TW Hya protoplanetary disk in visible total intensity scattered light. We have shown that the disk's azimuthal brightness asymmetries at r$>$40~AU, previously interpreted as arising from a misaligned inner disk and casting a single shadow, now appears to be comprised of two shadows, implying that a single disk at r$<$6~au is no longer a favored explanation as the originating shadowing structure. This is the first detection of a shadow splitting into more than one feature on a protoplanetary disk.

We instead consider two inclined rings at roughly the same orbital separation. This suggested explanation, when coupled with a series of gaps in dust density and moderate disk flaring matches both the one dimensional SED as well as the detailed features seen in scattered light. Nonetheless, the models presented here are not necessarily unique since it is beyond the scope of this paper to fully sample parameter space. It is true, however, that a detailed joint fit of the SED, resolved sub-mm continuum images, interferometry data, and scattered light in both total and polarized intensity represents a significant number of constraints on both the disk surface structure and the presence of interior tilted rings. The presence of shadows adds unique constraints to such modeling, hopefully in a way that provides a more unique model for TW Hya's disk.

The recent behavior of the shadow throws into question the shadow rotation period of 15.9~yr inferred by D17. This period was originally assumed to be a precession period of the inclined disk, which required a fairly massive companion to be consistent with observations. Additionally, the recent discovery of a similar shadow in sub-mm CO line emission \citep{teague22} suggests that instead of pure rigid rotation, the shadow(s) may oscillate in PA due to the mutual precession of the disks. Future observations will reveal whether the behavior of the shadow is truly periodic, and what that period relates to physically. If the shadow angular motions can be better disentangled, it could open up more powerful predictions for physical processes driving the shadow motions.

The presence of multiple tilted rings is highly suggestive of multiple planetary companions in the innermost regions of the TW~Hya disk. While planets have been searched for in NIR data at around 6~au by VB17, no candidates were found down to a few tens of Earth masses in the existing disk gaps, assuming the level of disk extinction and planet evolutionary state is well known. Additional unsuccessful planet searches have been conducted with the Keck Vector Vortex coronagraph \citep{ruane17} and in H~$\alpha$~\citep{cugno19,Follette22}. Intriguing compact sources in the sub-mm have also been pointed to as evidence for planetary companions in the outer disk \citep[e.g.][]{tsukagoshi19}. Additionally, the limited radial velocity data taken for TW~Hya in the NIR is not sufficient to test the prediction of giant planet mass companions at distances of 5-10~au by itself, and visible radial velocities are plagued by the significant accretion and stellar activity of TW Hya \citep{huelamo08}. The Gaia mission's DR3 astrometric residual noise is not unusual compared to other stars of similar magnitude. Limits from a combination of Hipparcos and Gaia EDR3 proper motion anomaly measures show that the motion is consistent with no observable acceleration and place limits of 1.32, 1.29, 1.30, and 6.69~M$_{\mathrm Jup}$ for companions at orbits of 3, 5, 10, and 30~au, suggesting that if any planets are causing the structures that they are likely less than these masses \citep{brandt21,kervella22}. Given the inferred orbital separation of the shadowing structures, the favorable face-on nature of the disk, and the sensitivity that the full Gaia astrometric observations will have, it is possible an astrometric signature of the perturbing companions could be detected in the future. The combination of a measured planet mass and resulting shadowing structure would mean that TW~Hya will remain a key object for the study of planet formation and disk-planet interactions.

\begin{acknowledgements}
Some of the data presented in this paper were obtained from the Mikulski Archive for Space Telescopes (MAST) at the Space Telescope Science Institute. The specific observations analyzed can be accessed via \dataset[10.17909/mtx2-vz53]{https://doi.org/10.17909/mtx2-vz53}. We thank the anonymous referee for their thoughtful comments. We additionally thank Emily Rickman for helpful discussions related to companion limits to TW Hya and Inbok Yeah for discussions related to MCFOST modeling. Support for program \#16228 was provided by NASA through a grant from the Space Telescope Science Institute, which is operated by the Association of Universities for Research in Astronomy, Inc., under NASA contract NAS5-26555. This work is also based on observations made with the NASA/ESA Hubble Space Telescope, obtained from the Data Archive at the Space Telescope Science Institute, which is operated by the Association of Universities for Research in Astronomy, Inc., under NASA contract NAS5-26555. These observations are associated with programs \#11608,13753,and 13665. The authors wish to thank the STScI Program Coordinator, S. Meyett, for her tireless work in getting the visits for Program 16228 scheduled successfully. RN acknowledges funding from UKRI/EPSRC through a Stephen Hawking Fellowship (EP/T017287/1). RA gratefully acknowledges funding from the European Research Council (ERC) under the European Union’s Horizon 2020 research and innovation programme (grant agreement No 681601), and from the Science \& Technology Facilities Council (STFC) through Consolidated Grant ST/W000857/1. This work has made use of data from the European Space Agency (ESA) mission
{\it Gaia} (\url{https://www.cosmos.esa.int/gaia}), processed by the {\it Gaia}
Data Processing and Analysis Consortium (DPAC,
\url{https://www.cosmos.esa.int/web/gaia/dpac/consortium}). Funding for the DPAC
has been provided by national institutions, in particular the institutions
participating in the {\it Gaia} Multilateral Agreement.

\end{acknowledgements}

\bibliographystyle{aasjournal}



\end{document}